\documentclass[review]{elsarticle}

\journal{Journal of Informetrics}

\biboptions{authoryear}

\usepackage{cleveref}
\usepackage{enumitem}
\usepackage{graphicx}
\usepackage{float}

\begin{document}

\begin{frontmatter}

\title{An empirical and theoretical critique of the Euclidean Index}

\author[mymainaddress]{Jens Peter Andersen\corref{mycorrespondingauthor}}
\ead{jpa@ps.au.dk}
\cortext[mycorrespondingauthor]{Corresponding author}

\address[mymainaddress]{Danish Centre for Studies on Research and Research Policy. Aarhus University. Bartholins Alle 7, Dk-8000 Aarhus C. Denmark.}

\begin{abstract}
The recently proposed Euclidean index offers a novel approach to measure the citation impact of academic authors, in particular as an alternative to the $h$-index. We test if the index provides new, robust information, not covered by existing bibliometric indicators, discuss the measurement scale and the degree of distinction between analytical units the index offers. We find that the Euclidean index does not outperform existing indicators on these topics and that the main application of the index would be solely for ranking, which is not seen as a recommended practice.
\end{abstract}


\end{frontmatter}


In a recent paper, \cite{Perry2016} propose an indicator of the impact of individuals, designed to provide a more rigorous approach hereto than the $h$-index which has been a de facto standard in several assessment situations, ranging from academic hiring to research grant applications. Their proposed indicator, the Euclidean index, is interpreted as the Euclidean length of the vector composed of the total citation scores of an individual's papers, or in other words the square root of the sum of squares of an author's citations per paper. While the authors claim this to be a new indicator, it is rather a modification of two previously proposed indicators; the $R$-index \citep{Jin2007} and the $e$-index \citep{Zhang2009}, which use the same general formulation as the Euclidean index, with only minor differences (further details in the methodology). Another variation of close resemblance is the Energy interpretation \cite{Prathap2011}. Despite this, the approach used by \citeauthor{Perry2016} warrants a discussion of this index in particular, and the properties of these types of indices in general.

\citeauthor{Perry2016} designed their index based on five axioms, which they consider crucial for an indicator of an individual's citation impact, and through two empirical tests, they find that it outperforms the $h$-index with regard to 1) maintaining inter-field rankings after rescaling and 2) matching a ranking of top universities created by the National Research Council (NRC) of the United States of America. The approach of using rigorous, pre-defined axioms for designing scientometric indices is commendable, and the resulting index possesses some qualities, which are obviously missing in the $h$-index, namely their five axioms; monotonicity, independence, depth relevance, scale invariance and directional consistency \citep{Perry2016}. However, we will claim that there are two essential discussions which are omitted in this paper, and a number of other papers related to the $h$-index; the validity and ethics of assessing individuals as well as the sufficiency of the proposed axioms. We will argue for the former in the following section, followed by a case for the latter, including macro-scale empirical analysis, in the remainder of the paper.

\section{Assessment of individuals}
All bibliometric research is divided into three parts, one of which is evaluative bibliometrics, descriptive bibliometrics another and referencing studies a third. Evaluative and descriptive bibliometrics may focus on citations or publications and are generally quantitative in nature. For both types of quantitative studies, methodological approaches depend on the aggregation level and thus the unit size being evaluated. Macro-scale studies focus on world- or nation-wide research outputs, meso-level look at large research units, such as universities and possibly university departments, while studies of individual authors are considered micro-scale bibliometric studies. It is essential for any type of quantitative bibliometric study to use these aggregated units of analysis just as well as citation counts are aggregated references, viewing not individual works but the collective works of larger units as the object of study. By doing so, we move from the sometimes arbitrary selections of one reference over another, due to norms, personal relations, availability etc., to a generalized, statistical view \citep{VanRaan1998a}. Whether these sociological, personal and normative variations at the reference level are actually randomly distributed over large aggregates is debatable \citep{Waltman2013b}, and certainly requires a considerable amount of care, delicacy and large numbers of publications per unit of analysis \cite  {Vinkler2007,Costas2010}. As a consequence, within the field of bibliometrics, analyses of individuals, or micro-level citation analysis, have focused on more descriptive elements, such as collaborative networks or factors influencing productivity \citep{Costas2010} or as a support tool for informed peer-review \citep{Aksnes2004}. Traditionally, ranking algorithms and hard impact assessments have not been applied to micro-level aggregates, such as authors, due to the sensitivity of indicators to even fairly small outliers when regarding small samples. Since the publication of the h-index \citep{Hirsch2005}, there has been a growing attention on the dangers of assessment on the individual level and the potential influence on the research process \citep{Vinkler2007,VanRaan2006a} but also a large number of papers utilizing and attempting to improve the h-index \citep{Bornmann2008,Braun2006,Moed2009b,Glaenzel2012}. \citeauthor{Perry2016} are also critical towards the concept of evaluating individuals through citation counts, and the $h$-index in particular, yet maintain that these assessments are a matter of fact:
\begin{quote}
	[Citation indices] are regularly used to inform critical decisions about funding, promotion and tenure. With decisions of this magnitude on the line, one should approach the problem of developing a good index as systematically as possible. \cite[2722]{Perry2016}
\end{quote}
We should certainly take into account those practices that are currently part of the norms of various scientific disciplines \citep{DeRijcke2015,Rushforth2015,DianaHicks2015}, and the types of indicators which are used and understood outside the sciento- and bibliometric fields, e.g. by funders, managers or scientists themselves \citep{Leydesdorff2016}. In assessing and developing new indicators, we should thus consider if they are tools for use of bibliometricians or end-users. The Euclidean index is clearly aimed at the latter, and it should therefore also be clearly understandable and interpretable by these.

\section{Assessment of indicators}

As mentioned above, \citeauthor{Perry2016} apply an axiomatic approach to indicator design, basing the justification of the Euclidean index on five, mathematical requisites for a bibliometric indicator of individual, scholarly impact. These axioms - monotonicity, independence, depth relevance, scale invariance and directional consistency - are common, mathematical requirements for good measurements, and the authors argue well for the properties of the Euclidean index and how the index performs better than the $h$-index at ranking economics scholars. However, the authors have not tested the performance of their indicator against basic bibliometric units, nor other existing indicators, and it is thus unclear whether this new indicator even offers new or more robust information. The only empirical test performed is on the ranking ability of the indicator versus the $h$-index, compared to a peer-produced ranking. As it has been recommended that ranking should not be performed on the individual level \citep{VanRaan2006a,Vinkler2007,Costas2010}, this practice should be avoided, and will not be pursued further in this paper, which will instead focus on the basic properties of all bibliometric indicators; size and impact. These two factors have been identified by \cite{Leydesdorff2009} and \cite{Bollen2009a} as the main categories in which the majority of bibliometric indicators can be arranged into. The $h$-index and its derivatives are most clearly separated from these two factors, although not necessarily for the better \citep{Leydesdorff2009,Bollen2009a,Waltman2012a,Perry2016}. The introduction of a new citation-based bibliometric indicator should provide new information or insight, in order to not be redundant. Alternatively, new indicators can seek to provide more precise or robust information on aspects of scientific outputs, thus replacing existing indicators. Both the novelty and robustness of the Euclidean index will be tested empirically and theoretically in this paper; in comparison with basic units as well as the $h$-, $R-$ and $e$-indices, which are the most closely related indicators.

Also the ordinality of the Euclidean index will be discussed, in particular with regard to distinguishing between two units of analysis. This is a problematic aspect of the index, as \citeauthor{Perry2016} claim the index can be used not only for ranking, but also for general statements about the underlying distributions, or at least comparisons between distributions. Given the definition of the index, and typical citation distributions, we consider this unlikely, and we will discuss this using a thought experiment and a simulation on empirical data. This is especially important, as \citeauthor{Perry2016} present their indicator as a standalone solution, while it has been recommended that bibliometric analyses should ideally operationalize multiple indicators \citep{DianaHicks2015}.

\subsection{Interpreting measurements}

As we claim in the introductory paragraphs, it is clear that the Euclidean index is designed with end-users (grant committees, administrators, researchers) in mind, and should therefore be easily interpretable in its own right. This can be achieved by operationalizing a unit \citep{Hand2004}, carrying an interpretational value either on the nominal, ordinal or scale level. Simple, well-known examples of this could be gender distinctions, US dollars and degrees Celsius, each of which have an intrinsic value. Perhaps the most interesting of these is the interval unit, degrees Celsius, as this scale is humanly defined and other scales such as Kelvin and degrees Fahrenheit are equally valid. However, by designing the scale around the freezing and boiling point of water, an additional, un-arbitrary value is added. With regard to bibliometric indicators, the same is true. Basic count variables such as number of publications ($P$) and citations ($C$) are their own units, similar to capita counts, but with e.g. average scores, this becomes less obvious. However, as we have the basic units $P$ and $C$, we can also express the unit of the mean as $C/P$ (see e.g. \cite{Vinkler2010}, also for a systematic approach to bibliometric measurement). Correspondingly for field-normalized citations, we can fairly easily interpret the resulting scores as a mean relative to the world (or database) average of a comparable paper. Also the $h$-index can be interpreted, albeit somewhat less clear and more arbitrarily \citep{Waltman2012a}, as the number $h$ of a ranked set, $C$, consisting of the citation counts per paper, in which $h$ is the highest rank equal to or higher than the corresponding citation score (see also Eq. \ref{eq:h}). For the Euclidean index, this interpretation is not the case, as the resulting indicator is on a scale with only one fixed point ($0$), and no intrinsic value of the unit. As an example, consider an author with two papers, of which paper $a$ is cited ten times and paper $b$ is cited just once. The sum of squares ($SS$) becomes $10^2+1^2 = 101$, and the Euclidean index is $10.05$, but we have no means of interpreting this particular number without further information. The combination of $SS$ and the square root hereof detaches the measurement from the original unit, $C$, and there is no replacement unit. In this, the Euclidean index somewhat resembles the $h$-index, which suffers from epistemologically parallel problems \citep{Waltman2012a}. Consequently, the $e$- and $R$-indices, relying on the $h$-index are also not readily interpretable. As a result, all of these indicators (and certainly a number of others) offer no value or context in their numerical representation, only allowing ordinal comparisons or rankings. This claim of ordinality differs slightly from the strict definition, as there are some interval- and-ratio like properties, however, not meeting the full requirements, we choose to refer to them as ordinal. This property is also further discussed in section \ref{sec:ordinality}. The opposite of this is illustrated by such indicators which have fixed, meaningful endpoints or centres from which other scores can be interpreted. Examples hereof are mean citation scores, mean normalized citation scores and percentile-based indicators (e.g. proportion of papers in the global top $5\%$).

Additionally, if paper $a$ is cited once more, the $SS$ becomes $(10+1)^2 + 1^2 = 122$, while if $b$ is cited instead of $a$, $SS = 10^2 + (1 + 1)^2 = 104$. While one might argue that this is a desirable effect and puts an emphasis on highly cited papers (as do \cite{Perry2016} with the inclusion of the \textit{depth relevance} axiom), it is also well-known that there is a Matthew effect leading to an overemphasis of papers that are already highly cited \citep{Small2004c}, e.g. through perfunctory referencing. It should thus not be necessary to artificially enhance the effect of proportionally few highly cited papers.

The remaining value of the index thus becomes a ranking of authors, which is of course considered necessary by some funders and employers, but as argued before, not a recommendable practice, due to the potential implications and built-in arbitrariness of any (micro-level) citation-based ranking.

\section{Empirical methodology}

In this part of the paper, we will first describe the basic, bibliometric indicators to which we compare the Euclidean index, followed by a description of the robustness test and finally the dataset used for the empirical analysis. 

\subsection{Assessed indicators}

With the aim of testing whether the Euclidean index provides new or more robust information we test its correlations with the two most basic measurements of size and impact; number of papers, $p$, and total number of citations, $c$. These two main dimensions have previously been used to test new indicators for their contributions \citep{Bollen2009a,Leydesdorff2009}. In addition, we test correlation with the parametric combination of the two, the mean number of citations per paper. In addition to these basic metrics, we also compare the Euclidean Index to the indicators: $h$-index \citep{Hirsch2005}, $R$-index \citep{Jin2007} and $e$-index \citep{Zhang2009}.

In the following, we will define the aforementioned metrics and indicators. Starting with the number of papers and sum of citations, we define the set of count of citations received by individual papers of a given author as $C$. Using this definition, the number of papers by this author is the cardinality of this set which we choose to denote $p$, and correspondingly, $c_i$ are the citations received by paper $i$ of $C$, which gives us the formal definition of total citations, $c$, for a given author (Eq. \ref{eq:c}) and the mean of the two (Eq. \ref{eq:mc}). We can here note that $C$ is geometrically related to the Euclidean index, in that the index is equivalent to the Euclidean distance from $A$ to $B$ described by the vector containing the citations in the set $C$. In this geometric interpretation, $c$ is equal to the Manhattan distance of the same vector. As a result, both indicators are expressions of Minkowski distances with different powers, and as such a direct relationship between the two should be expected.

\begin{equation}
c = \sum_{i=1}^{p}c_i
\label{eq:c}
\end{equation}

\begin{equation}
MC = \frac{c}{p}
\label{eq:mc}
\end{equation}

By introducing a ranking function $f(C)$ which ranks the distribution of $C$ according to $c_i$ in descending order, we can define the $h$-index as the maximal index, $i$, which at least corresponds to $c_i$:

\begin{equation}
h(C) = \max_i \min (f(C),i)
\label{eq:h}
\end{equation}

Several of the bibliometric indicators derived from the $h$-index utilize what is referred to as the $h$-core, namely the papers ranking up to $h$ in the above distribution. Also the $R$- and $e$-index focus on this core of publications, the former as the square root of the sum of citations (Eq. \ref{eq:R}), and the latter by subtracting the least possible citations for $h$, from this same distribution, thus focusing on the excess citations (hence the name $e$, Eq. \ref{eq:e}). By definition, the $h$-core will always contain $h$ publications, which will be cited at least $h$ times each. We therefore know that the $h$-core contains at least $h^2$ total citations, which can be seen in Eq. \ref{eq:e} as the main difference from the $R$-index. We also include a modification of the $R$-index proposed by \cite{Panaretos2009}, $R_m$ (Eq. \ref{eq:Rm}), which is essentially the same as the $R$-index, but individual citation scores are also transformed by their square root.

\begin{equation}
R = \sqrt{\sum_{i=1}^h{c_i}}
\label{eq:R}
\end{equation}

\begin{equation}
R_m = \sqrt{\sum_{i=1}^h{c_i}^{1/2}}
\label{eq:Rm}
\end{equation}

\begin{equation}
e = \sqrt{\left(\sum_{i=1}^h{c_i}\right)-h^2}
\label{eq:e}
\end{equation}

For good measure and comparison, we also document the Euclidean index here (Eq. \ref{eq:euc}), and as can be seen from the definition, we can expect a correlation with the total sum of citations, as $c_i$ is the independent variable in both definitions.

\begin{equation}
\iota_E = \sqrt{\sum_{i=1}^pc_i^2}
\label{eq:euc}
\end{equation}

It is obvious from formulations \cref{eq:R,eq:e,eq:euc} that there is an algebraic likeness of the $R$-, $e$- and $\iota_E$-indices. The main difference is the exclusion of papers cited less than $h$ times in the $R$- and $e$-indices and the squaring of citation counts in the $\iota_E$-index. We should thus expect a power function correlation, with some variance due to differences between the $h$-core and the full publication set. Regardless of such expected correlations, the foundation of these two indicators on the $h$-core directly implies that some of the axioms which the Euclidean index are designed around cannot be fulfilled. Both indices are \textit{monotonous}, but just like the $h$-index is not \textit{independent}, these indicators are not. They do however fulfill the requirement of \textit{depth relevance}, as both indicators operate on a maximization of the most highly cited documents. Following the argument of \citeauthor{Perry2016}, the $e$- and $R$-index both have the same problem with \textit{scale variance} as the $h$-index, and do not have this property. The \textit{directional consistency} proposed by \citeauthor{Perry2016} holds for both indicators.

Finally, we will also include the field normalized citation scores, using the approach of \cite{Waltman2012mncs}. As indicators, we will include both the total normalized citation score, $NCS$, of an author as well as the mean normalized citation score, $MNCS$. The formulations of these indexes are equivalent to those of $C$ and $MC$, but for field normalized citations.

 


\subsection{Robustness analysis}

Robustness, or stability, is an essential element of measurement. Robust indicators should provide the same value when performing a new measurement, and this interpretation makes sense in the experimental sciences from where the term originates. In fields such as bibliometrics where the sample usually is the (practical) population, repeating the experiment will always produce the same result. Bootstrapping can be a useful technique in this case, to create resamples based on the original sample/population, but by allowing for replacement, thus randomly omitting observations and repeating others for each resampling \citep{Efron1993a}. This also allows us to calculate the limits of the $95$ percent interval, corresponding to confidence intervals for actual sampling experiments. In bibliometrics, there seems to be a consensus on referring to these intervals as stability intervals, as they indicate how stable the reported indicator scores are, or how sensitive to outliers in the data \citep{Schneider2014}. The technique is problematic, as a) outliers in bibliometric data have a significant value, and should not be omitted and b) the population is the population\footnote{In this as previous mentions of "population", it should be interpreted as the practical population defined by the data available through the provider, in this case Thomson Reuters. This will of course differ from the real population, which we can never know.}, which defeats one major purpose of the resampling. Nonetheless, bootstrapping stability intervals provide a useful heuristic for assessing the sensitivity of a given observation. In combination with the correlation analyses, we can test both for novelty and increased robustness.

Bootstrapping was performed using the \verb|boot| package \citep{Canty2015a,Davison1997} for \verb|R version 3.2.3| \citep{RCoreTeam2015}. \verb|R| is also used for all other calculations and illustrations (using \verb|ggplot2| \citep{Wickham2009}).

\subsection{Citation data}

In order to provide realistic, empirical assessments of the Euclidean index, we use data on citation counts based on the Web of Science (WoS) Science Citation Index-Enhanced (SCI-E) and Social Science Citation Index (SSCI). However, we use a relational database version of these citation indices managed by the Centre for Science and Technology Studies at Leiden University (CWTS). Among other supplementary information, the database contains tables of author groupings, which are algorithmically generated links between unique authors and their papers \citep{Caron2014}. Like any approach to constructing complete author bibliographies, the method behind this author disambiguation is not complete, however, \citeauthor{Caron2014} report $95$ percent precision and $90$ percent recall per author, which is entirely sufficient for the present analysis. In addition, the algorithm generating these author profiles is conservative, so that new, potentially duplicate author profiles are created (even though they may be singletons) rather than assigning uncertain papers to existing authors. We therefore select only those authors who have written at least twenty papers. Subsequently we limit the publication set related to the authors to journal articles and reviews only, effectively removing letters, conference contributions etc. As a result, some authors will have fewer than twenty papers after this delineation. We further limit the dataset to only those authors with their first publication between 2000 and 2005. The intention of using these limitations is entirely to obtain a manageable set of authors, in this case resulting in $255,755$ observations. The publication years are chosen to reflect a current sample of publishing authors, while still allowing for some time to accrue citations. The selected authors have in total written $2,667,186$ papers until week 13, 2016. For all of these papers we gathered the total amount of citations up until the most recent update in 2016.

\section{Results}

In the following, we will first provide an overview of the relationship between the Euclidean index and the four other indicators, followed by a bootstrapping analysis, in combination to test for novelty. This is followed by a thought experiment and a simulation related to the ordinality and interpretation of the Euclidean index.

\subsection{Novelty and robustness}

\begin{figure}[H]
\caption{$\iota_E$ plotted as function of $P$, $C$, $MC$, $h$, $R$, $R_m$, $e$, $NCS$ and $MNCS$.}
\includegraphics[width=\textwidth]{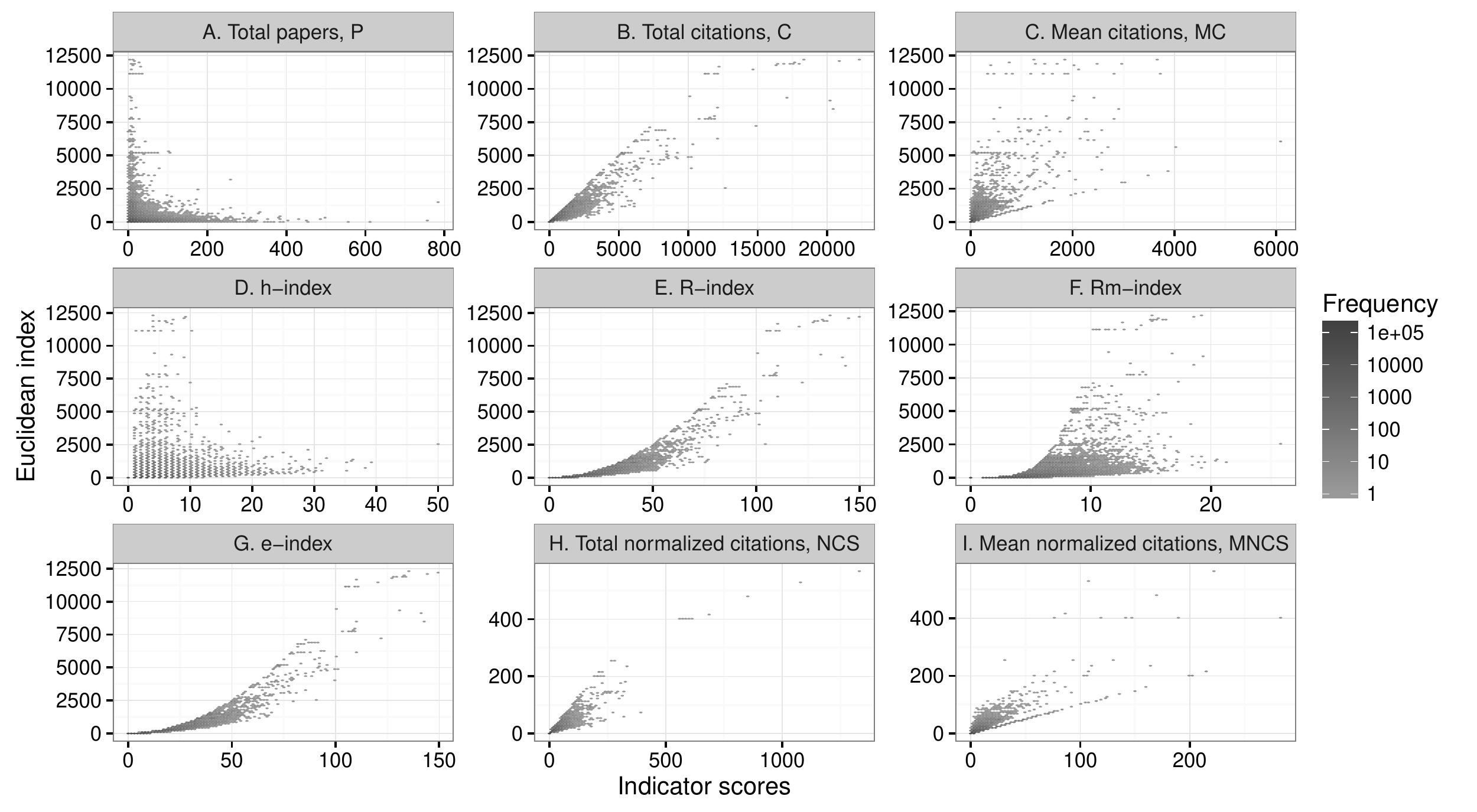}
\label{fig:correlations}
\end{figure}

In Figure \ref{fig:correlations} we plot the Euclidean index as a function of the six other indicators, and the two field-normalized versions. The most clear correlation for the basic units, as anticipated, is with the sum of citations per author, $C$. The total papers, $P$, at first glance are reciprocally correlated with $\iota_E$, but plotting $\iota_E$ against $P^{-1}$ does not provide a clear image of a linear relationship (plot not shown). Mean citations, $MC$, are not correlated with the Euclidean index, but interestingly, the function creates a lower and higher threshold for $\iota_E$ depending on $MC$, as is also the case for $C$. As also documented by \cite{Perry2016}, there is no correlation with the $h$-index. These relationships make a lot of sense, when considering the variables involved in each indicator, and the previous findings by \cite{Leydesdorff2009} and \cite{Bollen2009a}. The $e$- and $R$-indices are very clearly correlated with the Euclidean index, while it is more unclear for the two normalized indicators.

\begin{table}[H]
\caption{Pearson's $r$ for all pairs of indicators.}
\begin{tabular}{l|r r r r r r r r r}
\hline
 & $P$ & $MC$ & $h$ & $e$ & $R$ & $R_m$ & \textit{NCS} & \textit{MNCS} & $\iota_E$\\
\hline
$C$ & 0.32 & 0.91 & 0.87 & 0.99 & 0.99 & 0.99 & 0.87 & 0.83 & 0.99\\
$P$ &  & -0.03 & 0.46 & 0.28 & 0.31 & 0.38 & 0.21 & 0.29 & 0.27\\
$MC$ &  &  & 0.72 & 0.91 & 0.91 & 0.87 & 0.81 & 0.76 & 0.93\\
$h$ &  &  &  & 0.81 & 0.86 & 0.91 & 0.73 & 0.65 & 0.83\\
$e$ &  &  &  &  & 0.99 & 0.97 & 0.88 & 0.83 & 0.99\\
$R$ &  &  &  &  &  & 0.99 & 0.88 & 0.83 & 0.99\\
$R_m$ &  &  &  &  &  &  & 0.86 & 0.80 & 0.97\\
\textit{NCS} &  &  &  &  &  &  &  & 0.92 & 0.88\\
\textit{MNCS} &  &  &  &  &  &  &  &  & 0.84\\
\hline
\end{tabular}
\label{tab:pearson}
\end{table}

The inter-indicator correlations (all pairs) are analysed using Pearson's correlation coefficient, $r$. The results of these pairwise tests are listed in Table \ref{tab:pearson}. We can observe some degree of correlation for almost all indicator pairs, with the exception of $P$, which is correlated moderately with $h$, $R$ and $R_m$ only. $C$ on the other hand correlates strongly with all other indicators than $P$. These findings are in line with the aforementioned evidence of \textit{size} and \textit{impact} being the main dimensions of all indicators, except for the $h$-index. It is also clear from Table \ref{tab:pearson} that $\iota_E$ is aligned with the impact dimension, with the strongest correlations for $C$, $e$ and $R$. All the $h$-based indicators also have high pairwise correlations.

\begin{figure}[H]
\caption{Spearman's rank correlation coefficient for $\iota_E$ paired with $P$, $C$, $MC$, $h$, $R$, $R_m$, $e$, $NCS$ and $MNCS$.}
\includegraphics[width=\textwidth]{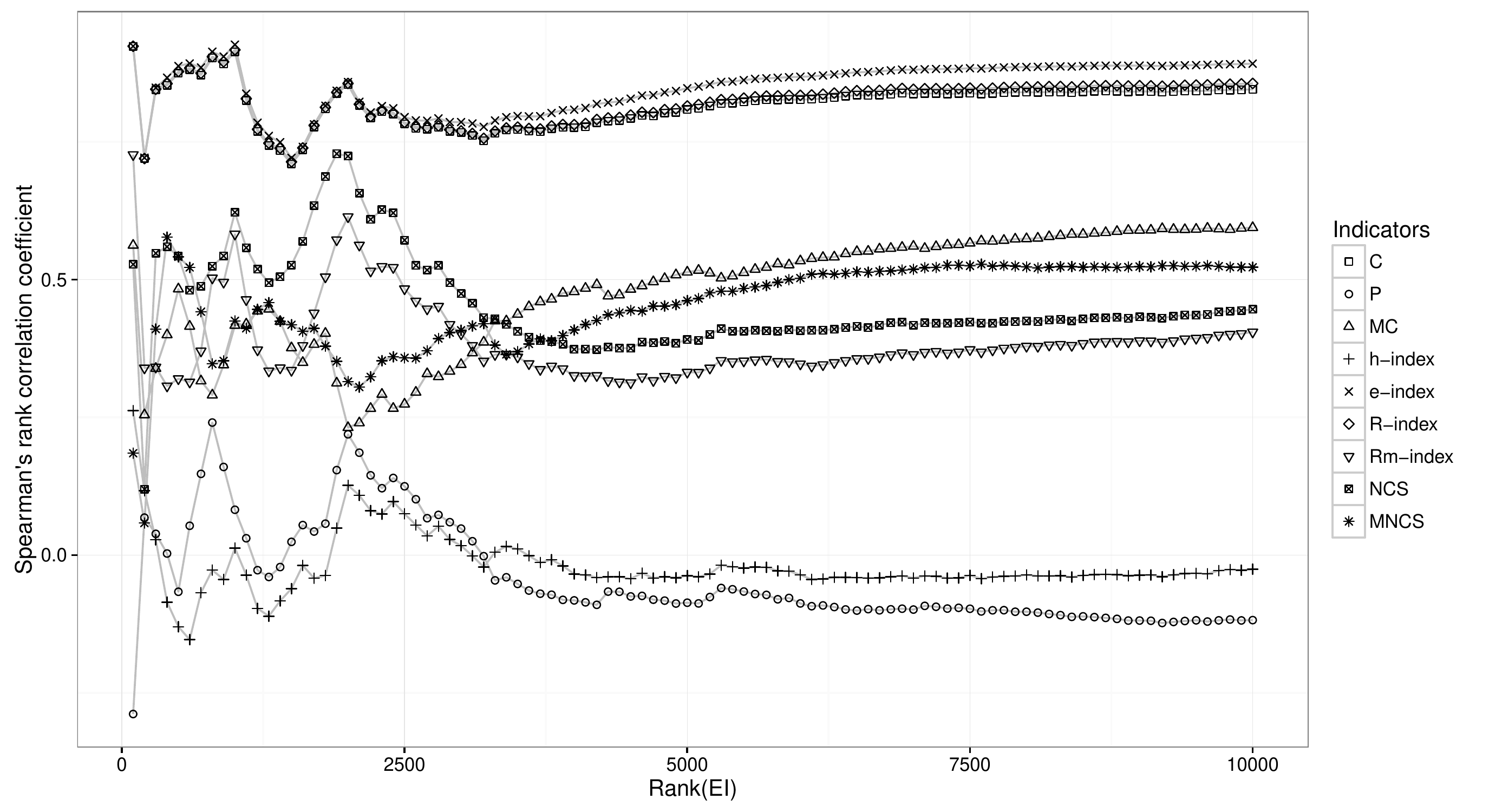}
\label{fig:spearman}
\end{figure}


Supplementing these absolute correlations, we also regard the ranking functions of the included indicators. While we have argued that rankings of individual academics should be avoided, it is nevertheless the aim of the Euclidean index, and a rank correlation analysis can assess whether the index indeed provides novel information on it's own terms. We use Spearman's rank correlation coefficient, $\rho$, to inform whether authors ranked by the Euclidean index are ranked similarly by other indicators. The results are provided in a brief overview in table \ref{tab:spearman} and as a function of sample size in Figure \ref{fig:spearman}. It is obvious from both analyses that the sum of citations, $C$, as well as the $e$-, $R$- and $R_m$-indices are most highly correlated with the Euclidean index. Several indicators actually are highly correlated with the Euclidean index when including all observations, however, this is biased by the discrete nature of many of these indicators and the underlying citation scores. We therefore included a limited set of observations, using the $10,000$ authors ranking highest by $\iota_E$. From table \ref{tab:spearman} we see the same three indicators correlating highly, while other indicators have more modest correlations, none at all ($h$-index) or even negative correlations ($P$). From Figure \ref{fig:spearman} we can confirm large fluctuations in these correlations when the sample size is more limited than $10,000$ observations. 

\begin{table}[htb]
\caption{Spearman's rank correlation coefficient, $\rho$, for correlation between $\iota_E$ and other indicators. Correlations are calculated for all authors as well as the $10,000$ top-ranked by $\iota_E$.}
\begin{tabular}{l r r}
& \multicolumn{2}{c}{Spearman's $\rho$}\\
Indicator & All observations & Top 10,000\\
\hline
$P$ & $.27$ & $-.120$\\
$C$ & $.99$ & $.850$\\
$MC$ & $.93$ & $.590$\\
$h$ & $.83$ & $-.026$\\
$e$ & $.99$ & $.890$\\
$R$ & $1.00$ & $.860$\\
$R_m$ & $0.97$ & $0.97$\\
$NCS$ & $.88$ & $.450$\\
$MNCS$ & $.84$ & $.520$\\
\hline
\end{tabular}
\label{tab:spearman}
\end{table}

\begin{figure}[H]
\includegraphics[width=\textwidth]{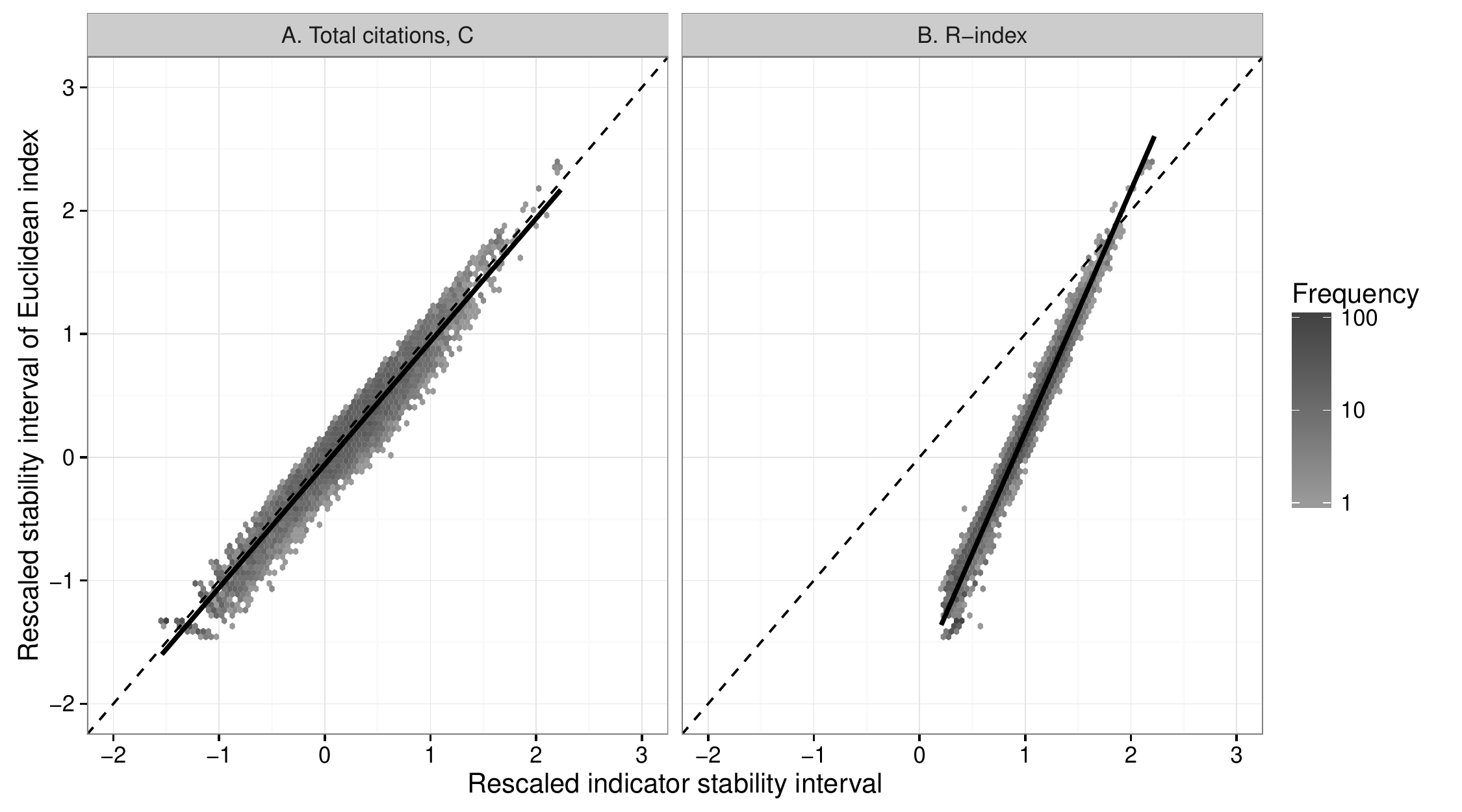}
\caption{Bootstrap stability intervals for $\iota_E$ and $C$. All observations have been transformed to $100 = \max$(indicator) and subsequently $\log$-transformed. Observations show the stability interval range of $\iota_E$ as a function of same for $C$ and $R$. The data is restricted to those authors with at least 50 publications and at least one citation. The dashed lines represents $y = x$, while the solid lines represent regressions on the correlations.}
\label{fig:bootstrap}
\end{figure}

This informs us that the Euclidean index does not provide us with new information, whether regarding absolute indicator scores or ranks, but rather transforms existing information. And while we have a rough idea about how to interpret a total citation count (although this is problematic as well), the Euclidean index is ordinal, and thus less informative, and the same can be said for the $e$- and $R$-indices. The question therefore is, whether the indicator is more robust than the total citation counts. We therefore bootstrap a subset of the authors, restricting it to those with more than 50 papers and at least one citation. This sample has $7,122$ observations, and for each observation resampling is performed with $1,000$ replications. We perform the bootstrapping on the Euclidean index, total citations and $R$-index, as these two indicators are both highly correlated with $\iota_E$ and with identical or less dispersed observations than $e$ and $R_m$. They should therefore be the most robust of the correlated indicators. In order to appropriately compare the indicators, which operate on different scales, we rescale them to an index $100 = \max$(indicator), and the same procedure is performed on the stability intervals of each bootstrapped paper set. We use $95$ percent stability intervals, and report the range of the interval in Figure \ref{fig:bootstrap}, from which it is clear that the sum of citations is just as stable an indicator as the Euclidean index, and the $R$-index is less stable. Linear regression strongly estimates the sum of citations and Euclidean index to have practically identical stability intervals ($R^2 = .956, \beta = .999$). The Euclidean index does thus not provide more robust information than the total citations, but clearly not less either. The index is clearly more stable than the $R$-index, especially in the lower range of stability intervals.

\subsection{Distinction and ordinality}
\label{sec:ordinality}

With regard to the interpretation of the Euclidean index as an ordinal index, which despite its ordinality can be used to determine to which degree an author's production is a multiple of another authors' \cite[2728-2729]{Perry2016}, we find it useful to illustrate with a thought experiment how this is strictly dependant on the definition and context of the assessment situation. We wish to investigate two perspectives of this: what is the potential difference in $\iota_E$ for two authors with equal total citations, and what is the potential difference in total citations for two authors with equal $\iota_E$. For the first perspective, assume an author with two papers = ${c_1 = 100, c_2 = 0}$ and another author with papers $a$ and $b$, where $c_a + c_b = 100$. We can then make a function of $c_b$ describing the $\iota_E$ of the second author:

\[\iota_E(c_b) = \sqrt{(100 - c_b)^2 + c_b^2}\]

\begin{figure}[htb]
\includegraphics[width=0.75\textwidth]{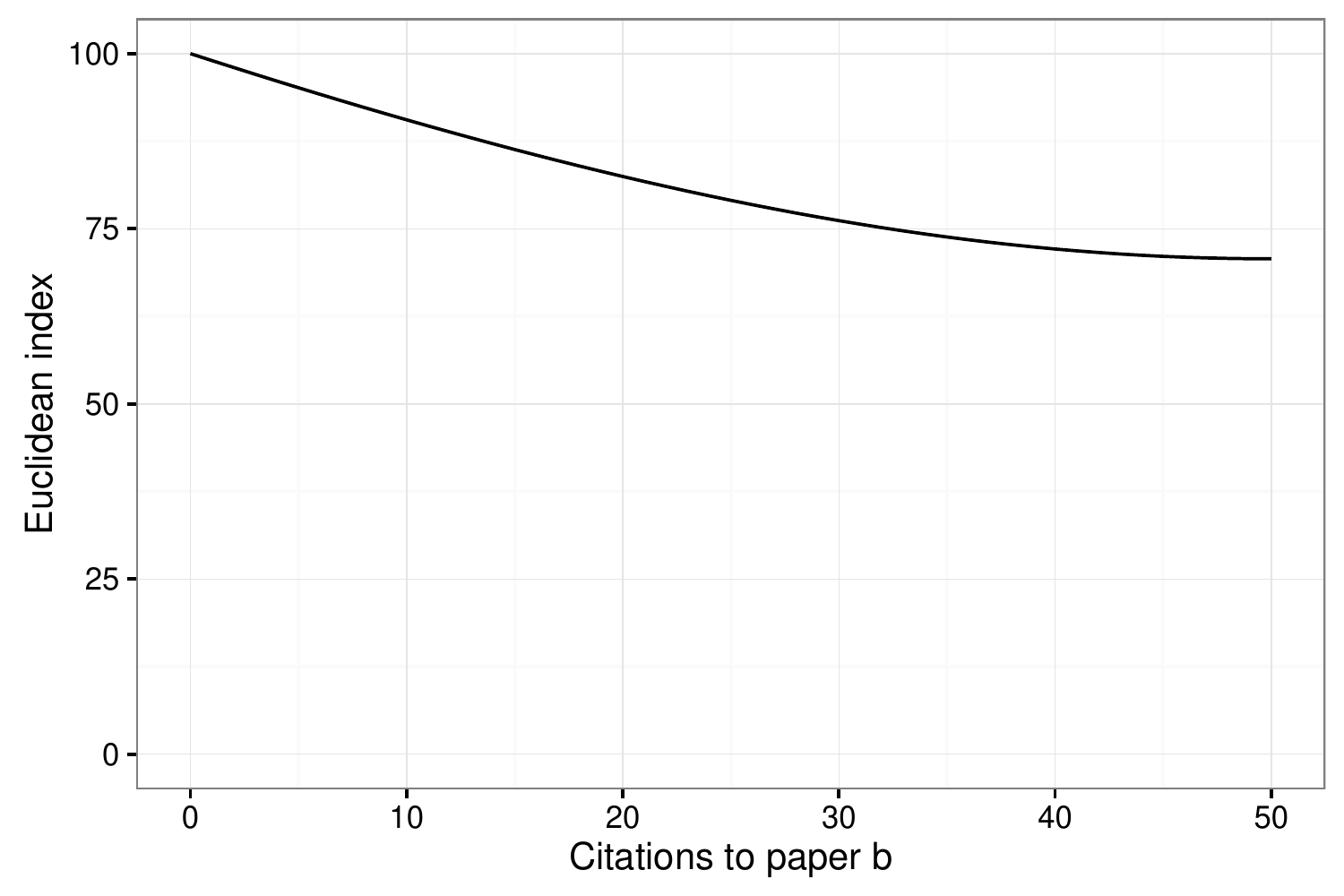}
\caption{Polynomial function for a two-paper model.}
\label{fig:twopaper_a}
\end{figure}

It is obvious that this is a polynomial function, and the accompanying distribution is shown in Figure \ref{fig:twopaper_a}.

\begin{figure}[htb]
\includegraphics[width=0.75\textwidth]{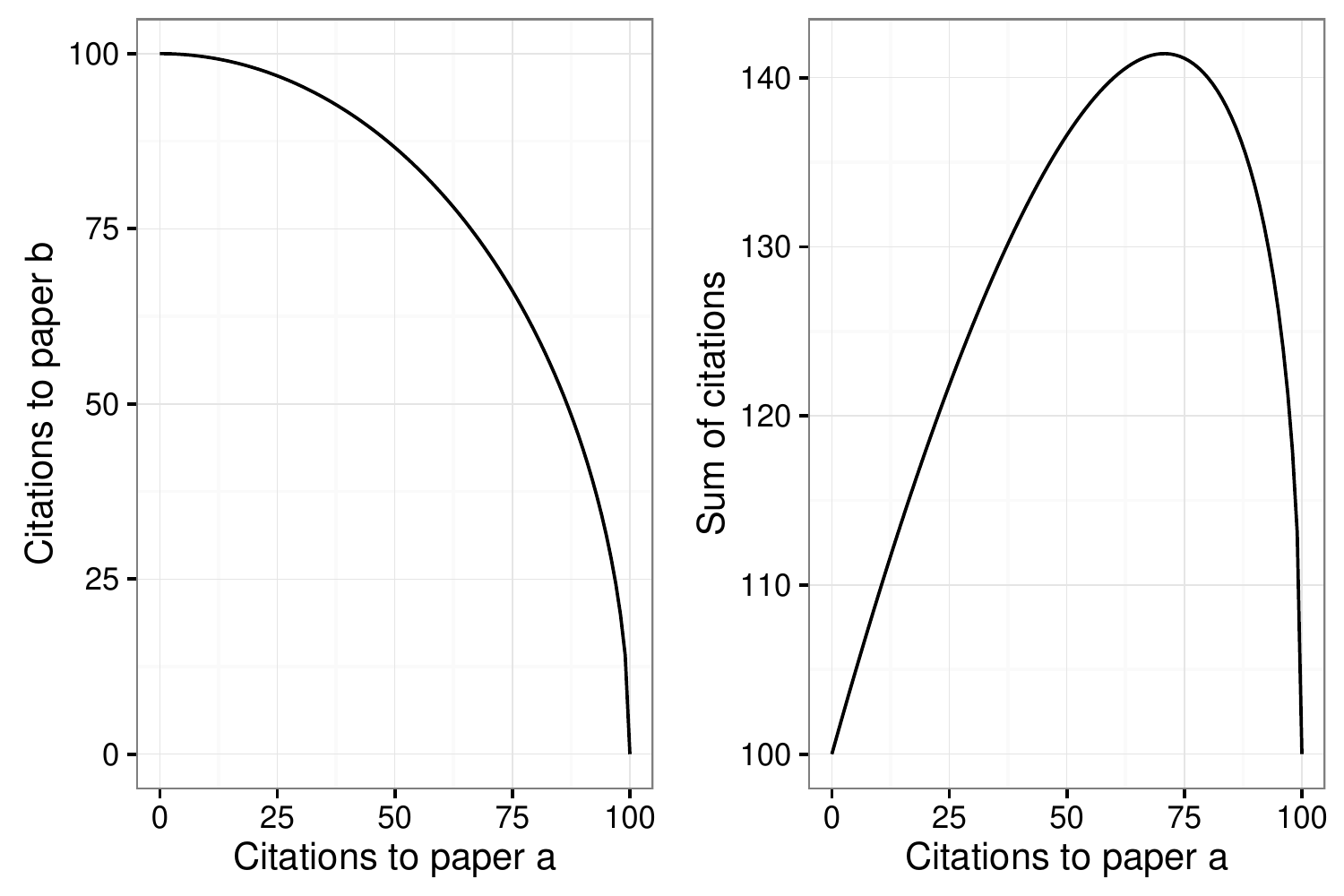}
\caption{Polynomial function for a two-paper model.}
\label{fig:twopaper_b}
\end{figure}

For the second perspective, we keep a constant $\iota_E = 100$ and using the same model before of an author with just two papers, we show here the function and distribution of this situation. We can describe the citations to $c_b$ as a function of $c_a$ and $\iota_E$ as \[c_b = \sqrt{\iota_E^2 - c_a^2}\] and subsequently a function \[f(c_a) = c_a + \sqrt{\iota_E^2 + c_a^2} = c_a + \sqrt{100^2 + c_a^2}\] showing the sum of citations to $a$ and $b$. The distribution of this function is shown in Figure \ref{fig:twopaper_b}

From these two thought experiments we can see that two authors with the same total citations can have large differences in $\iota_E$, with the largest differences found between authors who receive a steady amount of citations on each paper versus those with few exceptionally highly cited papers, a distinction also used in a previous analysis of the h-index, and referred to as "big producers" and "selective researchers" \citep{Costas2008}. We also see, that identical Euclidean indices for two authors can have quite different distributions behind them, and that statements such as \textit{"list $x$ is as good as the list that has as many papers as list $y$, but receives $\lambda$ times as many citations on each of them."} \cite[2729]{Perry2016} should be interpreted with extreme care, as the definition of "good" is entirely dependent on the assessment context. We therefore reaffirm that numerical representations of the Euclidean index are only ordinal, and do not have interval or ratio properties.
These observations of a very hypothetical situation directly imply consequences for real assessment situations, in which authors will benefit greatly from having single papers with very high citation scores. While this is actually part of the reasoning behind the Euclidean index, we claim that the underlying citation distributions are already so highly skewed that applying a square-transformation is not necessary, in order to assess the impact of individual authors. This is also related to the aforementioned Matthew-effect, and as a consequence hereof, the over-emphasis on more mechanistic, perfunctory citations to already highly cited papers as opposed to other types of potentially more relevant or meaningful citations. It is not our intention to argue against the use of indicators of citation impact, or the inclusion of outlier scores (which are actually quite relevant indeed), but merely against the artificial amplification of scores which are already sociologically amplified.

\begin{figure}
\includegraphics[width=0.9\textwidth]{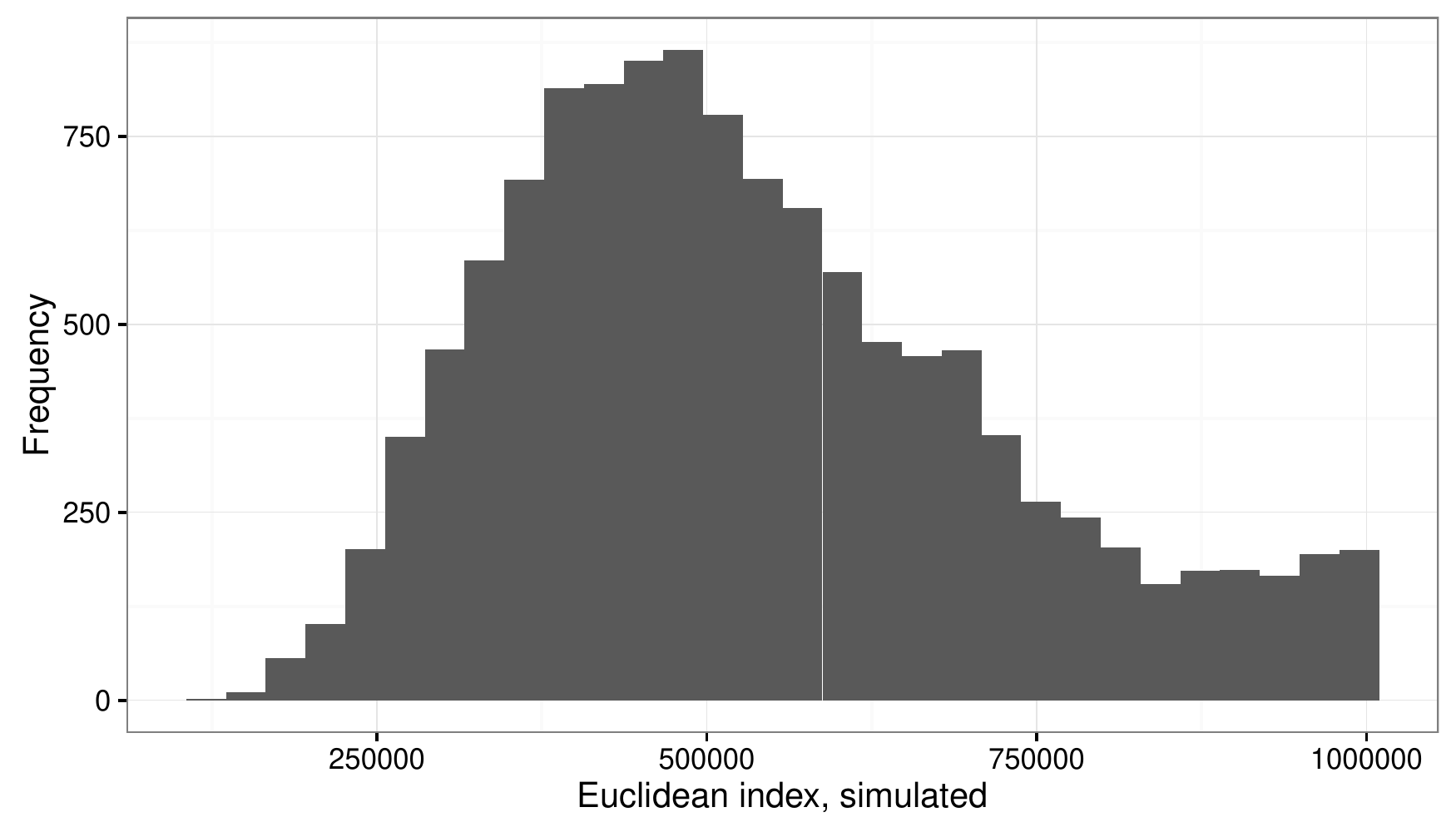}
\caption{Authors normalised to a mega-citation. Histogram showing the distribution of Euclidean index scores for all authors, if their entire set of papers was rescaled to receive one million citations in total. Limited to authors with $P >= 20$, $C >= 100$.}
\label{fig:eucsim }
\end{figure}

In a final thought experiment, we test further the degree to which the Euclidean index can be used to distinguish between two authors. By manipulating the actual citation distributions of our empirical set of authors, to all receive in total one million citations (but with the same relative distribution per author) we can compare a large set of realistically distributed authors with regard to the resulting Euclidean index. As such an analysis would be extremely sensitive to authors with just one cited paper, we limit our dataset to those with $P >= 20$ and $C >= 100$. A histogram showing the frequency of index scores is shown in Figure \ref{fig:eucsim }. We observe an almost Gaussian distribution of authors, showing us that if we observe a large number of authors with the same total citation score, a relatively small group of authors benefits greatly from the index. The main characteristic of these authors is a lower number of papers than others (data not shown here). The underlying effect of this is part of the very design of the indicator, in which citations to highly cited papers are weighted much higher (as a consequence of seeking to fulfill the depth relevance criterion).

\section{Discussion}

We have discussed and empirically analyzed the Euclidean index with respect to novelty, robustness, ability to distinguish between authors, ordinality and the interpretation of the index's scale. To our knowledge, this is the first large-scale application of the Euclidean index, and comparison to other indicators than the $h$-index. Our results show that the index is strongly correlated with the total citations for an author, and with similar stability intervals. The index thus does not contribute with new or more robust information, but a transformation of existing information. We can thus not recommend the use of this indicator over other, existing indicators. We have shown that authors with highly skewed citation distributions achieve higher scores on the Euclidean index than those with the same total citation count but less skewness. It is uncertain if this particular property could be applied in specific instances, but there is a clear risk of adverse effects if proper care is not taken, and we would ultimately recommend the use of a combined set of indicators, e.g. the $MNCS$ and the proportion of papers ranking in a given top percentile compared to the database, as these indicators are more informative, offering scale measurements rather than ranking schemes.

Through a thought experiment we have shown how the same Euclidean index score can be based on different distributions, and also the opposite is true. Even with extremely small samples such as the ones applied in the thought experiment, this can lead to a confusion as to the meaning of the score, and with more papers, the complexity increases. We have also shown that the index is dependent on which paper in a distribution receives a new citation, resulting in a situation in which some citations are more worth than other. Additionally, those papers which receive the greatest award from new citations are those which are highest cited already, thus reinforcing the Matthew effect already present in scholarly communication \citep{Merton1968,Small2004c}. These findings also raise questions about the consequences of the depth relevance criterion used in the design of the Euclidean index. The weak formulation used by \citeauthor{Perry2016}, that an indicator should not be maximized when citations are spread thinly is relevant and correct, but the resulting intentional maximization of highly cited papers is not. We also question the relevance of scale invariance, as there are well-established functions for field-, age- and type-normalization. Of the remaining axioms for indicator design, monotonicity is certainly relevant, but also fulfilled by any known indicator. The primary criteria allowing us to discern indicators are thus independence and directional consistency. As argued in this paper, as a result of the missing independence of the $h$-index \citep{Waltman2012a,Perry2016}, indicators derived from the $h$-property are also not independent, including in this case the $e$-, $R$ and $R_m$-indices.

Finally we have argued that the scale of the Euclidean index is arbitrary and thus of limited use to those it seems intended for; the research managers, funding organizations and scientists themselves. The main function of the index is as a ranking mechanism. While this also seems to be the primary intent of \citeauthor{Perry2016}, they also claim a scalable kind of ordinality. While this is to some extent true, it is highly dependent on the assessment situation. This is of course true for all indicators of scholarly impact, and as argued above, assessment should never rely on (one) bibliometric indicator(s) alone, especially in the case of micro-level analysis \citep{Aksnes2004,DianaHicks2015,Leydesdorff2016,Vinkler2007,Waltman2013b,Costas2010}. Should the Euclidean index be applied for ranking, our analysis shows evidence that the index provides the same kind of information as the $e$-, $R$- and $R_m$-indices, with better stability intervals than these. The Euclidean index does present a viable alternative in this case, but one should keep the emphasis on highly cited authors in mind. While all four indices emphasize these, we can see from the formulations and distributions that the $R_m$-index is the more conservative, while the Euclidean index applies the strongest weight on the highest cited papers.

There are several useful and sensible other approaches to micro-level bibliometric analyses, and there are indicators which can provide supporting information to peer-reviewing of individuals \citep{Aksnes2004}, but as has been stated before, and as the results of our study as well as that of \citeauthor{Perry2016} shows, ranking individuals by any indicator for assessment leads to arbitrary selections based on purely statistical attributes of a biased indicator\footnote{All indicators distill information, and through this bias certain elements.} based on imperfect data. While we support designing indicators not just for bibliometricians, but also managers and scientists etc., we also strongly oppose the concept of ranking individuals, and recommend that indicators should be designed for more than ordinality.

\section{Acknowledgments}

The author wishes to thank Jesper W. Schneider and Carter W. Bloch for helpful comments and discussion, as well as three anonymous and very helpful referees.

\section*{References}

\bibliography{mybibfile}

\end{document}